\documentclass[aps,prd,superscriptaddress,nofootinbib]{revtex4}

\usepackage{epsfig}
\usepackage{bm}
\usepackage{amssymb}
\usepackage{amsmath}
\usepackage{color}
\usepackage{subfigure}
\usepackage[colorlinks,
            linkcolor=blue,
            anchorcolor=black,
            citecolor=blue
            ]{hyperref}

\newcommand{\bx}{\boldsymbol{x}}
\newcommand{\bb}{\boldsymbol{b}}
\newcommand{\br}{\boldsymbol{r}}
\newcommand{\bp}{\boldsymbol{p}}

\newcommand{\bk}{\boldsymbol{k}}

\newcommand{\rmd}{\mathrm{d}}

\newcommand{\be}{\begin{equation}}
\newcommand{\ee}{\end{equation}}
\newcommand{\bea}{\begin{eqnarray}}
\newcommand{\eea}{\end{eqnarray}}

\allowdisplaybreaks[4]

\begin{document}
\title{Two particle azimuthal harmonics in pA collisions}

\author{Manyika Kabuswa Davy}
\affiliation{Key Laboratory of Quark and Lepton Physics (MOE) and Institute
of Particle Physics, Central China Normal University, Wuhan 430079, China}

\author{Cyrille Marquet}
\affiliation{CPHT, \'Ecole Polytechnique, CNRS, Universit\'e Paris-Saclay, F-91128 Palaiseau, France}

\author{Yu Shi}
\affiliation{Key Laboratory of Quark and Lepton Physics (MOE) and Institute
of Particle Physics, Central China Normal University, Wuhan 430079, China}

\author{Bo-Wen Xiao}
\affiliation{Key Laboratory of Quark and Lepton Physics (MOE) and Institute
of Particle Physics, Central China Normal University, Wuhan 430079, China}
\affiliation{CPHT, \'Ecole Polytechnique, CNRS, Universit\'e Paris-Saclay, F-91128 Palaiseau, France}

\author{Cheng Zhang}
\affiliation{Key Laboratory of Quark and Lepton Physics (MOE) and Institute
of Particle Physics, Central China Normal University, Wuhan 430079, China}

\begin{abstract}
We compute two-particle production in $pA$ collisions and extract azimuthal harmonics, using the dilute-dense formalism in the Color Glass Condensate framework. The multiple scatterings of the partons inside the projectile proton on the dense gluons inside the target nucleus are expressed in terms of Wilson lines. They generate interesting correlations, which can be partly responsible for the signals of collectivity measured at RHIC and at the LHC. Most notably, while gluon Wilson loops yield vanishing odd harmonics, quark Wilson loops can generate sizable odd harmonics for two particle correlations. By taking both quark and gluon channels into account, we find that the overall second and third harmonics lie rather close to the recent PHENIX data at RHIC.
\end{abstract}
%\pacs{24.85.+p, 12.38.Bx, 12.39.St, 12.38.Cy}
\maketitle

\section{Introduction}

The collectivity phenomenon in small systems, which is manifested in terms of multiple particle azimuthal correlations in $pp$ and $pA$ collisions \cite{Khachatryan:2010gv, CMS:2012qk, Abelev:2012ola, Aad:2012gla, Adare:2013piz, Adare:2014keg, Khachatryan:2015waa, Aidala:2018mcw}, has become an interesting topic of great importance in heavy ion physics. It is characterized by the anisotropic distribution of particles measured in the final state of very high multiplicity events, which can be decomposed into Fourier harmonics as $v_n \equiv \langle \cos n \Delta \phi \rangle$,
%follows
%\begin{equation}
%v_n \equiv \langle \cos n \Delta \phi \rangle,
%\end{equation}
where $\Delta \phi$ is the azimuthal angle difference between the measured particle and the reference particle or the reaction plane. 
On the one hand, it seems that the relativistic hydrodynamics approach, which provides the collective description of quarks and gluons after they are created, can describe RHIC and the LHC data very well\cite{Habich:2014jna, Shen:2016zpp}, despite the conventional belief that $pA$ and $pp$ systems are too small to form quark-gluon plasma medium. On the other hand, in the Color Glass Condensate (CGC) framework, initial state effects in $pA$ collisions can also generate collectivity among final-state particles, due to the multiple interactions between the projectile proton and dense gluons inside the heavy nuclear target \cite{Armesto:2006bv, Dumitru:2008wn, Gavin:2008ev, Dumitru:2010mv, Dumitru:2010iy, Kovner:2010xk, Kovchegov:2012nd, Dusling:2012iga, Dumitru:2014dra, Dumitru:2014yza, Dumitru:2014vka, Lappi:2015vha, Schenke:2015aqa, Lappi:2015vta,McLerran:2016snu, Kovner:2016jfp, Iancu:2017fzn, Dusling:2017dqg, Dusling:2017aot, Fukushima:2017mko, Kovchegov:2018jun, Boer:2018vdi, Mace:2018vwq, Mace:2018yvl,Kovchegov:2013ewa,Altinoluk:2018ogz,Kovner:2018fxj}.

Usually, it is convenient to use the so-called dilute dense factorization in the CGC framework to compute observables in $pA$ collisions. In this approach, the proton projectile is relatively dilute as compared to the target nucleus, which allows us to approximately neglect any multiple scattering between the spectators inside the proton projectile and active partons. In very high multiplicity events, it is natural to assume that there can be a few active partons from the proton projectile participating the interaction with the target nucleus \cite{Lappi:2015vha, Lappi:2015vta, Dusling:2017dqg, Dusling:2017aot}. As far as two particle correlations are concerned, we can take two independent partons (quarks or gluons) from the proton side and compute the multiple scatterings of these two partons with the target nucleus. The leading $N_c$ contribution of these type of multiple scattering can be written as two independent dipole amplitudes, and therefore keep these two incoming partons independent and generate no correlation.

Furthermore, there are also interesting sub-leading $N_c$ contributions, which arise by breaking color neutral dipoles and converting them into color quadrupoles or even higher correlators in the course of multiple gluon exchanges with the target nucleus. As we shall see below in detail both analytically and numerically, these sub-leading $N_c$ contributions indeed can be responsible for the azimuthal harmonics in $pA$ collisions. In addition, when one or two of these two partons is gluon, one can find that the correlations are completely even with vanishing odd harmonics. In other word, in this simple model, the odd harmonics, such as $v_3$ and $v_5$, can only come from the case that the two incoming partons are two quarks. This is in agreement with the results in Ref.~\cite{Kovchegov:2012nd, Dusling:2017dqg}, which pointed out that the corresponding two-gluon productions only have even harmonics. Also, to obtain odd harmonics with incoming gluon states, one has to consider much more sophisticated interactions proposed in Ref.~\cite{Schenke:2015aqa, McLerran:2016snu, Kovner:2016jfp, Kovchegov:2018jun, Mace:2018vwq, Mace:2018yvl}.

The objective of this paper is to study the anisotropic harmonics within the aforementioned simple model in the dilute-dense factorization at lowest order, while the calculation considered in Ref.~\cite{Kovchegov:2018jun} is regarded higher order in $\alpha_s$. Our approach is closely related to the model calculation proposed in Ref.~\cite{Lappi:2015vha, Lappi:2015vta, Dusling:2017dqg, Dusling:2017aot} with a few minor differences. Generally speaking, as compared to earlier studies, we consider all possible combination of incoming partons in terms of quark and gluon degrees of freedom when we compute the azimuthal harmonics, and we focus more on the analytical understanding of these harmonics of particle correlations in the color dipole interpretation. 

The paper is organized as follows. In Sec.II, we briefly introduce our dilute-dense formalism for particle production and comment on the physical reason why quarks can generate odd harmonics as opposed to vanishing odd harmonics in gluon production. In Sec. III, both even and odd Fourier harmonics are derived for quark-quark, quark-gluon and gluon-gluon channels. As a conclusion, the phenomenological implication of our results are discussed in Sec. IV.
%Before we conclude in Sec. V, the phenomenological implication of our results are discussed in Sec. IV.  

\section{Dilute-dense framework for two-particle correlations}

Let us first recall the dilute-dense factorization framework frequently used to compute single-inclusive production in $pA$ collisions, also known as the hydrid factorization formula \cite{Dumitru:2002qt, Chirilli:2011km}. Denoting the transverse momentum $k_\perp$ and the rapidity $y$, the parton-level production cross-section can be written as 
\be
\frac{d\sigma}{dy d^2k_\perp} = x_pq(x_p) \left< a_q(k_\perp) \right>_{x_A}+ x_pg(x_p) \left< a_g(k_\perp) \right>_{x_A}
\label{hybrid}
\ee
with $q(x_p)$ (resp. $g(x_p)$) the collinear quark (reps. gluon) density inside the projectile proton, $x_p=\frac{k_\perp}{\sqrt{s}}e^y$, $x_A=\frac{k_\perp}{\sqrt{s}}e^{-y}$ and
\bea
a_q(k_\perp)&=&\int \frac{d^2{\bf x} d^2{\bf y}}{(2\pi)^2}\ e^{i k_\perp\cdot ({\bf x}-{\bf y})} \frac{1}{N_c} {\text tr} \left(V({\bf x})V^\dagger({\bf y})\right)
= \frac{1}{N_c} {\text tr} \left| \int \frac{d^2{\bf x}}{2\pi}\ e^{i k_\perp\cdot {\bf x}}\ V({\bf x}) \right|^2\ ,\\
a_g(k_\perp)&=&\int \frac{d^2{\bf x} d^2{\bf y}}{(2\pi)^2}\ e^{i k_\perp\cdot ({\bf x}-{\bf y})} \frac{1}{N_c^2\!-\!1} {\text Tr} \left(U({\bf x})U^\dagger({\bf y})\right)
= \frac{1}{N_c^2\!-\!1} {\text Tr} \left| \int \frac{d^2{\bf x}}{2\pi}\ e^{i k_\perp\cdot {\bf x}}\ U({\bf x}) \right|^2\ .
 \label{sq}
\eea
Here $\left< a_{q,g}(k_\perp) \right>_{x_A}$ indicates the color averaging of the fundamental ($V$) and adjoint ($U$) Wilson lines (yielding fundamental and adjoint Wilson loops, or color dipoles) in the gluon background fields of the target nucleus. The expectation value of the amplitude $\left< a_{q,g}(k_\perp) \right>_{x_A}$ essentially provides the transverse momentum $k_\perp$ of the order of the so-called saturation momentum $Q_s$. We shall perform those target averages using the McLerran-Venugopalan (MV) model \cite{McLerran:1993ni,McLerran:1993ka}.

Before the color average takes place, it is important to note that, in general
\be
a_q(-k_\perp)=\int \frac{d^2{\bf x} d^2{\bf y}}{(2\pi)^2}\ e^{i k_\perp\cdot ({\bf x}-{\bf y})} \frac{1}{N_c} {\text tr} \left(V({\bf y})V^\dagger({\bf x})\right)
\neq a_q(k_\perp). 
\ee
This is because for fundamental Wilson lines, prior to the average over the color configuration of the target, one has:
\be
{\text tr} \left(V({\bf y})V^\dagger({\bf x})\right)=\left[{\text tr} \left(V({\bf x})V^\dagger({\bf y})\right)\right]^*\neq
{\text tr} \left(V({\bf x})V^\dagger({\bf y})\right)\ .
\ee
which is equivalent to say that $2i\ {\text Im}\ {\text tr} \left(V({\bf x})V^\dagger({\bf y})\right) ={\text tr} \left(V({\bf x})V^\dagger({\bf y})\right)- {\text tr} \left(V({\bf y})V^\dagger({\bf x})\right) \neq 0$. Even though $a_q(k_\perp)$ and $a_q(-k_\perp)$ are real (since they can be written as squares as shown in Eq.~\ref{sq}), that non-zero imaginary part contributes to them with different signs. It is the target averaging which puts this imaginary part to zero\footnote{We stick here to the original MV model with a quadratic weight function. A non-zero $ \left<{\text Im} \,{\text tr} \left(V({\bf x})V^\dagger({\bf y})\right)\right>_{x_A}$ can be obtained with cubic terms, see e.g. \cite{Lappi:2016gqe}} for single quark production: $\left<{\text tr} \left(V({\bf x})V^\dagger({\bf y})\right)\right>_{x_A}=\left<{\text tr} \left(V({\bf y})V^\dagger({\bf x})\right)\right>_{x_A}$, and therefore we do have $\left< a_q(k_\perp) \right>_{x_A}=\left< a_q(-k_\perp) \right>_{x_A}$.
For gluons however, due to the fact that the adjoint representation is real, one has ${\text Im}\ {\text Tr} \left(U({\bf y})U^\dagger({\bf x})\right)=0$ and $a_g(k_t) = a_g(-k_t)$ configuration-by-configuration, as noticed in Ref.~\cite{McLerran:2016snu}. This difference between quarks and gluons has important consequences when looking at two-particle production, as we sketch now, prior to making more detailed calculations in the next Section.

Note first that we do not consider here the so-called jet contributions, which involve a single parton coming from the projectile that then splits into two, and which have been discussed extensively in several works \cite{Marquet:2007vb,Albacete:2010pg,Stasto:2011ru,Lappi:2012nh}. Indeed, those ``jet" contributions are subtracted from data prior to the extraction of the azimuthal harmonics we set out to calculate. For the sake of the argument, we can write in our dilute-dense approach the two-gluon production cross-section as (the proper impact-parameter treatment will be restored later):
\be
\frac{d\sigma}{dy_2 d^2k_1 dy_2 d^2k_2} = (x_1+x_2) g(x_1,x_2) \left< a_g(k_1) a_g(k_2) \right>_{x_A}\
\label{doublehybrid}
\ee
where $ (x_1+x_2) g(x_1,x_2)$ denotes a collinear double-gluon distribution, $x_i=\frac{k_i}{\sqrt{s}}e^{y_i}$, and $x_A=x_1\ e^{-2y_1}+x_2\ e^{-2y_2}$. In the same way that the gluon part of formula \eqref{hybrid} can be obtained from the dilute-dense $k_t$-factorization formula for single-inclusive gluon production after taking the collinear limit for the dilute projectile, formula \eqref{doublehybrid} can be obtained from the "squared" contribution of the $k_t$-factorization formula for two-gluon production given in \cite{Kovchegov:2013ewa} (see formula (65) there). Due to the fact that $a_g(k_2) = a_g(-k_2)$, it can generate no odd harmonics, as explained in \cite{McLerran:2016snu}. However, it does generate even harmonics (as we show below) and is not irrelevant for correlations, contrary to statements made in the literature. 

The other part of the formula in \cite{Kovchegov:2013ewa} (see formula (70)), the "crossed" contribution, would turn into a hybrid formula involving a double generalized distribution (with gluons having different transverse coordinates in the amplitude and the conjugate amplitude) on the projectile side, along with a quadrupole target average. We shall not include such a contribution in our model. It does not generate odd harmonics either, as the quadrupole terms are $k_2\to -k_2$ symmetric. It would be however interesting to see what happens in the hybrid limit to this crossed contribution, which is responsible for the so-called HBT terms and Bose enhancement contributions to even harmonics \cite{Altinoluk:2018ogz}.

Instead, what we add in our approach are quarks. The two-quark production cross-section mirrors eq.~\eqref{doublehybrid}, and corresponds to what was considered in \cite{Dusling:2017dqg,Dusling:2017aot} where only quarks were present: 
\be
\frac{d\sigma}{dy_2 d^2k_1 dy_2 d^2k_2} = (x_1+x_2) q(x_1,x_2)  \left< a_q(k_1) a_q(k_2) \right>_{x_A}\ .
\label{hybridvalence}
\ee
That brings us to the main point of this introductory section. As we noted before, prior to target averaging, $a_q(k_1) a_q(k_2)\neq a_q(k_1) a_q(-k_2)$. But what is different now with respect to the single-inclusive case, is that even after the MV model averaging, we still have
$\left< a_q(k_1) a_q(k_2) \right>_{x_A} \neq \left<a_q(k_1) a_q(-k_2)\right>_{x_A}$.
This is due to the fact that
$\left< {\text tr} \left(V({\bf x})V^\dagger({\bf y})\right) {\text tr} \left(V({\bf x'})V^\dagger({\bf y'})\right) \right>_{x_A}\neq
\left< {\text tr} \left(V({\bf x})V^\dagger({\bf y})\right) {\text tr} \left(V({\bf y'})V^\dagger({\bf x'})\right) \right>_{x_A}$.
Indeed,
\bea
\left< {\text tr} \left(V({\bf x})V^\dagger({\bf y})\right) {\text tr} \left(V({\bf x'})V^\dagger({\bf y'})\right) \right>_{x_A}-
\left< {\text tr} \left(V({\bf x})V^\dagger({\bf y})\right) {\text tr} \left(V({\bf y'})V^\dagger({\bf x'})\right) \right>_{x_A}=\label{originofv3}\\
2i\left< \left[ {\text Re}\,{\text tr} \left(V({\bf x})V^\dagger({\bf y})\right) +i\ {\text Im}\, {\text tr} \left(V({\bf x})V^\dagger({\bf y})\right) \right]  {\text Im} \, {\text tr} \left(V({\bf x'})V^\dagger({\bf y'})\right) \right>_{x_A}=\nonumber\\
-2\left< {\text Im}\, {\text tr} \left(V({\bf x})V^\dagger({\bf y})\right){\text Im} \,{\text tr} \left(V({\bf x'})V^\dagger({\bf y'})\right) \right>_{x_A}\neq 0\ .\nonumber
\eea
As we show in detail below, this can also be seen explicitly from the large-$N_c$ expansion (the full result was computed in Ref.~\cite{Dominguez:2008aa}, but we shall only need the first $1/N_c^2$ correction shown here):
\be
\left< {\text tr} \left(V({\bf x})V^\dagger({\bf y})\right) {\text tr} \left(V({\bf x'})V^\dagger({\bf y'})\right) \right>_{x_A}=
\left< {\text tr} \left(V({\bf x})V^\dagger({\bf y})\right)\right>_{x_A} \left< {\text tr} \left(V({\bf x'})V^\dagger({\bf y'})\right) \right>_{x_A}
+\frac{1}{N_c^2} \Delta(\bf x, \bf y, \bf x^\prime, \bf y^\prime)+\cdots.
\ee
The first term, which represents two independent dipole amplitudes, does not contribute to the difference \eqref{originofv3}, but the second term does: $\Delta({\bf x}, \bf y, \bf x^\prime, {\bf y^\prime})\neq$ $\Delta({\bf x}, \bf y, \bf y^\prime, \bf x^\prime)$.

Since $\Delta(\bf x, \bf y, \bf x^\prime, \bf y^\prime)$, which is explicitly given below, contains odd powers of $(\bf x -\bf y).(\bf x^\prime -\bf y^\prime)$, as pointed out in Ref.~\cite{Lappi:2015vha, Lappi:2015vta,  Dusling:2017dqg, Dusling:2017aot}, the higher-order $N_c$ corrections of the expectation value of the two-dipole amplitude generate finite value for all the odd Fourier harmonics, which involve $\left< a_q(k_1) a_q(k_2) - a_q(k_1) a_q(-k_2)\right>_{x_A}$ (multiple scattering must be included though, as in the dilute-dilute limit, odds harmonics are indeed absent: $\Delta_{dilute}\sim[({\bf x} -{\bf y}).({\bf x}^\prime -{\bf y}^\prime)]^2$). In the dipole model language, the correlation is generated due to the transition between two dipoles and the quadrupole as shown in Fig.~\ref{2dipole}, and the imaginary part discussed above arises due to an odd number of gluon exchanges. This cannot happen with gluon dipoles, this is why the two-gluon contribution \eqref{doublehybrid}, as well as the gluon-quark contribution which we also include, generate even harmonics only ($\left< a_g(k_1) a_{q,g}(k_2) \right>_{x_A} = \left<a_g(k_1) a_{q,g}(-k_2)\right>_{x_A}$). Furthermore, in the dilute-dense CGC calculation of sea-quark correlations \cite{Altinoluk:2016vax}, odd harmonics will a priori be generated by the very same two-fundamental-dipole correlators which we consider in our simplified model, but in that case the contribution from quarks and antiquarks will cancel each other since $a_{\bar{q}}(k)=a_q(-k)$ and so $a_{q}(k)+a_{\bar{q}}(k)=a_q(-k)+a_{\bar{q}}(-k)$. Therefore, when calculating odd harmonics from \eqref{hybridvalence}, only valence quarks will contribute. When calculating even harmonics, the sign of $k_{1,2}$ does not matter so the model applies two quarks, two antiquarks, or a mixture of both.

Finally, when we compute the Fourier harmonics from two-particle correlations, we follow Refs.~\cite{Lappi:2015vha, Lappi:2015vta,  Dusling:2017dqg, Dusling:2017aot} and use Wigner distributions for the projectile,
\begin{equation}
W({\bf b}, p_\perp) =\frac{1}{\pi^2} e^{-{\bf b}^2/{B_p}} \frac{\Delta^2}{4\pi}\int d^2 r_\perp e^{-ip_\perp\cdot r_\perp} e^{-\frac{1}{4}\Delta^2 r_\perp^2}= \frac{1}{\pi^2} e^{-{\bf b}^2/{B_p}-p_\perp^2/\Delta^2}\ ,
\end{equation}
instead of the collinear double-parton distribution for the incoming partons. This form assumes Gaussian distributions for both the impact parameter $\bf b$ and the transverse momentum $p_\perp$ with the variances $B_p$ and $\Delta^2$, respectively. This allows to restore the impact-parameter dependencies, but also to study the effect that a small parton transverse momentum on the projectile side would have. One could use $B_p \Delta^2 =1$, but instead we choose to treat these two parameters independently. This will allow us to vary separately the projectile transverse area $B_p$ and the average intrinsic transverse momentum $\Delta$. The product $B_p \Delta^2$ can be viewed as number density of the corresponding parton. 

\section{Azimuthal harmonics}

\subsection{Two-quark production}

\begin{figure}[t]%[tbp]
%\vskip0.0\linewidth
\begin{center}
\includegraphics[width = .95\linewidth]{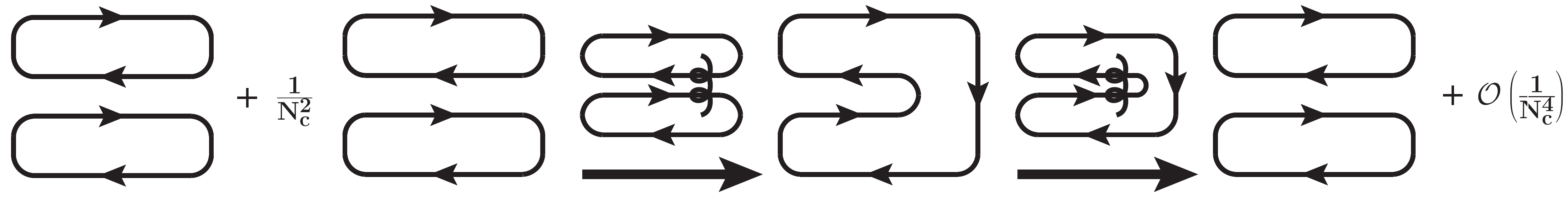}
\end{center}
\caption[*]{The two-dipole amplitude $\left<D\right>\left<D\right>$ plus the $\Delta/N_c^2$ correction arising from the transition to quadrupole configuration and then back to the two-dipole configuration.}
\label{2dipole}
\end{figure}

Let us first study the case of two incoming quarks from the projectile proton. The production rate is sensitive to the color averaging of two-dipole amplitude. By definition, the 2-dipole correlator, which is vital for the generation of two-particle correlation, can be written as
\begin{eqnarray}
\left<D\left(\bx_1,\bx_2\right)D\left(\bx_3,\bx_4\right)\right>\equiv\frac{1}{N_c^2}\left<\text{tr}\left[V(\bx_1)V(\bx_2)^{\dagger}\right]\text{tr}\left[V(\bx_3)V(\bx_4)^{\dagger}\right]\right>.
\end{eqnarray}
%which is the same result in \cite{Xie:2015gdj}, and we will use the functional form in \cite{Dominguez:2008aa}
By changing coordinates $\bx_{1,2}=\bb_1\pm\frac{\br_1}{2},\ \ \bx_{3,4}=\bb_2\pm\frac{\br_2}{2}$, 
%\begin{eqnarray}
%\bx_{1,2}=\bb_1\pm\frac{\br_1}{2},\ \ \ \bx_{3,4}=\bb_2\pm\frac{\br_2}{2},\ \ ...\label{br}
%\end{eqnarray}
and using the same matrix technique developed and commonly used in CGC\cite{Gelis:2001da,Fujii:2002vh,Blaizot:2004wv, Dominguez:2008aa, Dominguez:2012ad,Shi:2017gcq}, we can find (we neglect the dipole-size logarithmic dependence of $Q_s$):
\begin{eqnarray}
&&\left< D\Big(\bb_1+\frac{\br_1}{2},\bb_1-\frac{\br_1}{2}\Big)D\Big(\bb_2+\frac{\br_2}{2},\bb_2-\frac{\br_2}{2}\Big)\right>\Bigg|_{\text{to the}\frac{1}{N_c^2}\text{order}}\nonumber
\\
&=&e^{-\frac{Q_s^2}{4}(r_1^2+r_2^2)}\left[1+\frac{(\frac{Q_s^2}{2}\br_1\cdot \br_2)^2}{N_c^2}  \int_0^1 d\xi \int_0^{\xi} d\eta
e^{\frac{\eta Q_s^2}{8} [(\br_1-\br_2)^2-4(\bb_1-\bb_2)^2]} \right],
\end{eqnarray}
where $\bb_i$ and $\br_i$ is the central coordinate and the radius of the $i^{th}$ dipole respectively. The above two terms bear clear physical interpretation in the dipole formalism as illustrated in Fig.~\ref{2dipole}. The first term is simply from the scattering amplitudes of two independent dipoles on the target nucleus, while the second term represents the contribution due to color transitions from two-dipole to a quadrupole\cite{JalilianMarian:2004da, Dominguez:2011wm} and then back to two-dipole configuration\cite{Xie:2015gdj}. For such a color transition, the resulting amplitude is suppressed by a factor of $\frac{1}{N_c}$. In the meantime, one gains correlations between these two color dipoles by breaking them into color quadrupoles. In fact, the integrated variables $\xi$ and $\eta$ indicate the location of the above two color transitions from the front to the back of the target nucleus. 

The 2-particle anisotropic flows and cumulants \cite{Dusling:2017aot,Borghini:2001vi,Dumitru:2014yza,Khachatryan:2015waa} are defined by
\begin{eqnarray}
v_n\{2\} \equiv \sqrt{c_n\{2\}},
\ \ \
c_n\{2\} \equiv \left< e^{in(\phi_1-\phi_2)} \right> \equiv \frac{\kappa_{n}\{2\}}{\kappa_{0}\{2\}},
\end{eqnarray}
where $\kappa_{n}\{2\}$ is the $n^{\text{th}}$ harmonic distribution of the $2$-particle production 
\begin{eqnarray}
\kappa_n\{2\} \equiv \int \prod_{i=1}^2 \rmd^2\bp_i e^{ in(\phi_1  - \phi_{2}) } \frac{\rmd^2 N}{\prod_{i=1}^2 \rmd^2\bp_i},
\end{eqnarray}
where $\frac{\rmd^2N}{\prod_{i=1}^2 \rmd^2\bp_i}$ is the $2$-quark inclusive spectra %, which can be written as
\begin{eqnarray}
\frac{\rmd^{2}N}{ \rmd^2\bp_1 \rmd^2\bp_2 }
&=&\prod_{i=1}^2 \int \frac{\rmd^2\bb_i \rmd^2\br_i}{\Delta^2 B_p} \frac{\rmd^2\bk_i}{(2\pi)^2} W(\bb_i,\bk_i) e^{i(\bp_i-\bk_i) \cdot \br_i } \left< \prod_{j=1}^2 D \left(\bb_j+\frac{\br_j}{2},\bb_j-\frac{\br_j}{2}\right) \right>\nonumber
\\
&=&\prod_{i=1}^2 \int \frac{ \rmd^2\bb_i \rmd^2\br_i }{4\pi^3B_p} e^{ -\frac{b_i^2}{B_p} - \frac{\Delta^2r_i^2}{4}
+ i\bp_i \cdot \br_i } \left< \prod_{j=1}^2 D \left(\bb_j+\frac{\br_j}{2},\bb_j-\frac{\br_j}{2}\right) \right>,
\end{eqnarray}
where we used the Gaussian type Wigner function $W(b,k)\equiv\frac{1}{\pi^2}e^{-\frac{b^2}{B_p}-\frac{k^2}{\Delta^2}}$. It is straightforward to compute $\kappa_0\{2\}$ in this model, and find that it is normalized to unity as follows
\be
\kappa_0\{2\} =\int \rmd^2\bp_1 \rmd^2\bp_2 \frac{\rmd^2N}{\rmd^2\bp_1 \rmd^2\bp_2}=1.\label{k0}
\ee
%\begin{eqnarray}
%W(b,k)\equiv\frac{1}{\pi^2}e^{-\frac{b^2}{B_p}-\frac{k^2}{\Delta^2}}.
%\end{eqnarray}
%For the 2-quark case, the $0^{\text{th}}$ harmonic of the inclusive spectra is normalized to 
%\begin{eqnarray}
%\kappa_0\{2\}
%&=&\int \rmd^2\bp_1 \rmd^2\bp_2 \frac{\rmd^2N}{\rmd^2\bp_1 \rmd^2\bp_2}\nonumber\\
%&=&\prod_{i=1}^2 \int \frac{ \rmd^2\bb_i \rmd^2\br_i \rmd^2\bp_i }{4\pi^3B_p} e^{ -\frac{b_i^2}{B_p} - \frac{\Delta^2+Q_s^2}{4}r_i^2 + i\bp_i\cdot\br_i } \left[ 1 + \frac{ ( \frac{Q_s^2}{2} \br_1 \cdot \br_2 )^2 }{N_c^2} \int_0^1 \rmd\xi \int_0^{\xi} \rmd\eta e^{ \frac{\eta Q_s^2}{8} (\Delta\br_{12}^2 - 4\Delta\bb_{12}^2) } \right]\nonumber\\
%&=&\prod_{i=1}^2 \int \frac{ \rmd^2\bb_i \rmd^2\br_i }{4\pi^3B_p}e^{-\frac{b_i^2}{B_p}} (2\pi)^2 \delta^2(\br_i)\nonumber\\
%&=&1,\label{k0}
%\end{eqnarray}
Using the short hand notation $\Delta \bx_{ij}\equiv \bx_i-\bx_j$, the integrated $\kappa_{2n}\{2\}$ can be computed step by step as follows
\begin{eqnarray}
\kappa_{2n}\{2\}
&=&\int \rmd^2\bp_1 \rmd^2\bp_2 e^{ i2n\Delta\phi_{12} } \frac{\rmd^2 N}{ \rmd^2\bp_1 \rmd^2\bp_2 }\nonumber
\\
&=&\prod_{i=1}^2 \int \frac{ \rmd^2\bb_i \rmd^2\br_i \rmd^2\bp_i }{4\pi^3B_p} e^{ -\frac{b_i^2}{B_p} - \frac{\Delta^2+Q_s^2}{4}r_i^2 + i\bp_i\cdot\br_i } \left[ 1+ \frac{ (\frac{Q_s^2}{2}\br_1\cdot\br_2)^2 }{N_c^2} \int_0^1 \rmd\xi \int_0^{\xi} \rmd\eta e^{ \frac{\eta Q_s^2}{8} (\Delta\br_{12}^2-4\Delta\bb_{12}^2) } \right] e^{ i2n\Delta\phi_{12} }. \,\,\,%\nonumber
\end{eqnarray}
We first integrate over $\phi_i$ (the azimuthal angle of the $i^{\text{th}}$ outgoing quark momentum $\bp_i$) by using Eq. (\ref{intphi}), and obtain
\begin{eqnarray}
\kappa_{2n}\{2\}
=
\int_0^1 \rmd\xi \int_0^{\xi} \rmd\eta \prod_{i=1}^2 \int \frac{ \rmd^2\bb_i \rmd^2\br_i p_i\rmd p_i }{2\pi^2B_p} e^{ -\frac{b_i^2}{B_p} - \frac{\Delta^2+Q_s^2-\frac{\eta}{2} Q_s^2}{4} r_i^2 } J_{2n}(p_ir_i) \frac{ ( \frac{Q_s^2}{2} \br_1\cdot\br_2 )^2 }{N_c^2} e^{ \frac{\eta Q_s^2}{8} (-2\br_1\cdot\br_2 - 4\Delta\bb_{12}^2) } e^{ i2n\Delta\theta_{12} }, \nonumber
\end{eqnarray}
where $\theta_i$ is the azimuthal angle of the $i^{\text{th}}$ dipole size $\br_i$. Then by integrating over $\bb_i$ with the help of Eq. (\ref{intb}), and integrating over $\theta_i$ with Eq. (\ref{inttheta}) after expanding $e^{-\frac{\eta Q_s^2}{4}\br_1\cdot\br_2}$ into Taylor series, we arrive at
\begin{eqnarray}
\kappa_{2n}\{2\}
&=&\int_0^1 \rmd\xi \int_0^{\xi} \rmd\eta \prod_{i=1}^2 \int \frac{ \rmd^2\br_i p_i\rmd p_i }{2\pi^2B_p} e^{ -\frac{\Delta^2+Q_s^2-\frac{\eta}{2} Q_s^2}{4} r_i^2 } J_{2n}(p_ir_i) \frac{\pi^2B_p^2}{1+\eta Q_s^2B_p} \frac{ ( \frac{Q_s^2}{2} r_1 r_2 \text{cos} \Delta\theta_{12} )^2 }{N_c^2}\nonumber
\\
&&\times \sum_{l=0}^\infty \frac{ ( -\frac{\eta Q_s^2}{4} r_1 r_2 \text{cos} \Delta\theta_{12} )^{l} }{l!} e^{ i2n\Delta\theta_{12} }. \\
%\end{eqnarray}
%\begin{eqnarray}
%\kappa_{2n}\{2\}
&=&\int_0^1 \rmd\xi \int_0^{\xi} \rmd\eta \prod_{i=1}^2 \int \frac{ r_i\rmd r_i p_i\rmd p_i }{2\pi} e^{ -\frac{\Delta^2+Q_s^2-\frac{\eta}{2} Q_s^2}{4} r_i^2 } J_{2n}(p_ir_i) \frac{1}{1+\eta Q_s^2B_p} \frac{4}{N_c^2\eta^2}\nonumber
\\
&&\times \sum_{m=n-1}^\infty \frac{ ( -\frac{\eta Q_s^2}{4} r_1 r_2 )^{2m+2} }{(2m)!} \frac{ (2\pi)^2 \binom{2m+2}{m+n+1} }{2^{2m+2}},\label{k2nrp}
\end{eqnarray}
It is important to note that all the odd power of $( r_1 r_2 \text{cos} \Delta\theta_{12} )^{l}$ vanishes after $\theta_i$ integrations, therefore, we can set $l=2m$ for the computation of even harmonics. As to the odd harmonics, similarly, we can set $l$ to $2m+1$. Therefore, we can conclude that odd harmonics come from the odd power of $(\br_1\cdot \br_2)$ terms inside the double dipole expectation value, which is quite useful for us to show that the corresponding gluon productions yield no odd harmonics. 

Finally, by integrating over $r_i,\ p_i$ up to $\infty$ with the help of Eq. (\ref{intrp}), we can cast $\kappa_{2n}\{2\}$ into 
\begin{eqnarray}
\kappa_{2n}\{2\}
=
\frac{4}{N_c^2} \int_0^1 \rmd\xi \int_0^{\xi} \rmd\eta \frac{1}{1+\eta Q_s^2B_p} \frac{n^2}{\eta^2} \sum_{m=n-1}^\infty \frac{ (m!)^2 \binom{2m+2}{m+n+1} }{(2m)!} \left( \frac{\eta Q_s^2}{8a_q } \right)^{2m+2},\ \ \ a_q=\frac{\Delta^2+Q_s^2-\frac{\eta}{2} Q_s^2}{4}.\label{k2n}
\end{eqnarray}
Similar derivations can be applied to the $(2n+1)^{\text{th}}$ moment, which allows one to obtain 
\begin{eqnarray}
\kappa_{2n+1}\{2\}
=
\frac{4}{N_c^2} \int_0^1 \rmd\xi \int_0^{\xi} \rmd\eta \frac{1}{1+\eta Q_s^2B_p} \frac{(n+\frac{1}{2})^2}{\eta^2} \sum_{m=n-1}^\infty
\frac{ [(m+\frac{1}{2})!]^2 \binom{2m+3}{m+n+2} }{(2m+1)!} \left(  \frac{\eta Q_s^2}{8a_q } \right)^{2m+3}.\label{k2n+1}
\end{eqnarray}

The asymptotic behavior of the integrated harmonics in the quark-quark channel can be computed analytically. In the small $Q_s$ limit, we find
\begin{eqnarray}
\kappa_{n}\{2\} \overset{\text{small }Q_s}{\simeq}
%\begin{cases}
%\frac{\pi}{N_c^2} \left(\frac{Q_s^2}{4\Delta^2}\right)^3 \left( 1-\frac{9Q_s^2}{4\Delta^2}-\frac{Q_s^2B_p}{2}\right), &n=1,
%\\
%\\
\frac{4}{N_c^2} \frac{(\frac{n}{2}!)^2}{n!}\left(\frac{Q_s^2}{2\Delta^2}\right)^{n} \left[ 1 - \left(\frac{n}{2} + 1 - \frac{1}{n+1}\right)\frac{Q_s^2}{\Delta^2} -\left(1-\frac{2}{n+1}\right)Q_s^2B_p \right]
,  &n\geq2,
%\end{cases}
\label{knsmallQ}
\end{eqnarray}
and in large $Q_s$ regions, we get
\begin{eqnarray}
\kappa_{n}\{2\} \overset{\text{large }Q_s}{\simeq}
\begin{cases}
%\frac{1}{N_c^2} (-\frac{1}{3}+\frac{\pi}{4}) \frac{1}{Q_s^2B_p}, &n=1,
%\\
\frac{1}{N_c^2} \left[ -\frac{1}{4} + \ln2 + \ln(Q_s^2B_p) \right] \frac{1}{Q_s^2B_p}, &n=2,
\\
\frac{1}{N_c^2} \frac{n^2}{(n-2)(n+2)} \frac{1}{Q_s^2B_p}, &n\geq3.
\end{cases}\label{knlargeQ}
\end{eqnarray}
We have also compared our results with the numerical results in Ref.~\cite{Dusling:2017aot} after setting $\Delta^2B_p=1$, and the numerical values for $v_n$ roughly agree. The differences can be understood as a result of the large $N_c$ approximation that we employ in this calculation as well as the difference in the range of $\bp$ and $\br$ integrations. 

%For a special case of the 2-quark azimuthal correlation, we substitute $n=1$ into Eq. (\ref{k2n}), which gives the the $2^{\text{th}}$ harmonic
%\begin{eqnarray}
%\kappa_{2}\{2\}=
%\frac{8}{N_c^2} \int_0^1 \rmd\xi \int_0^{\xi} \rmd\eta \frac{1}{1+\eta Q_s^2B_p} \frac{1}{\eta^2} \left[ 1 + \frac{2}{1-A^2} + \frac{3\ln(1-A^2)}{A^2} \right]
%\Bigg|_{ A = \frac{\eta Q_s^2}{ 8a_q } },
%\end{eqnarray}
%by which we plot $v_2\{2\}$ and $c_2\{2\}$ depending on $Q_s^2$ and $Q_s^2B_p$ in Fig.\ref{v22c22}, and the latter one shows a falloff approximately proportional to $\frac{\ln Q_s^2B_p}{Q_s^2B_p}$ in large $Q_s^2B_p$ region, which has been referred in \cite{Dusling:2017aot}.

\begin{figure}[t]
\vskip0.0\linewidth
\centerline{
\includegraphics[width = 0.45\linewidth]{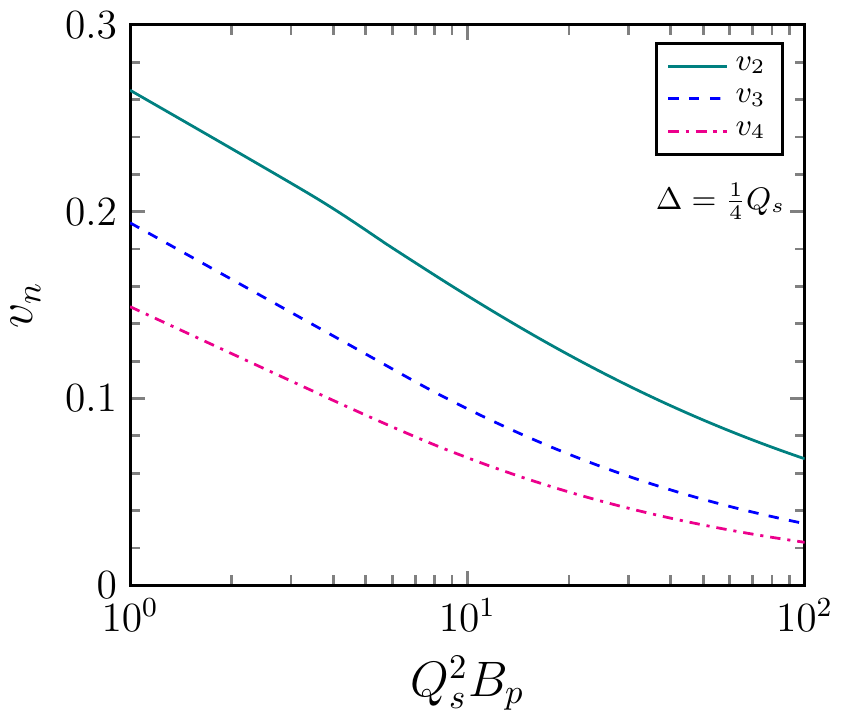}\includegraphics[width = 0.45\linewidth]{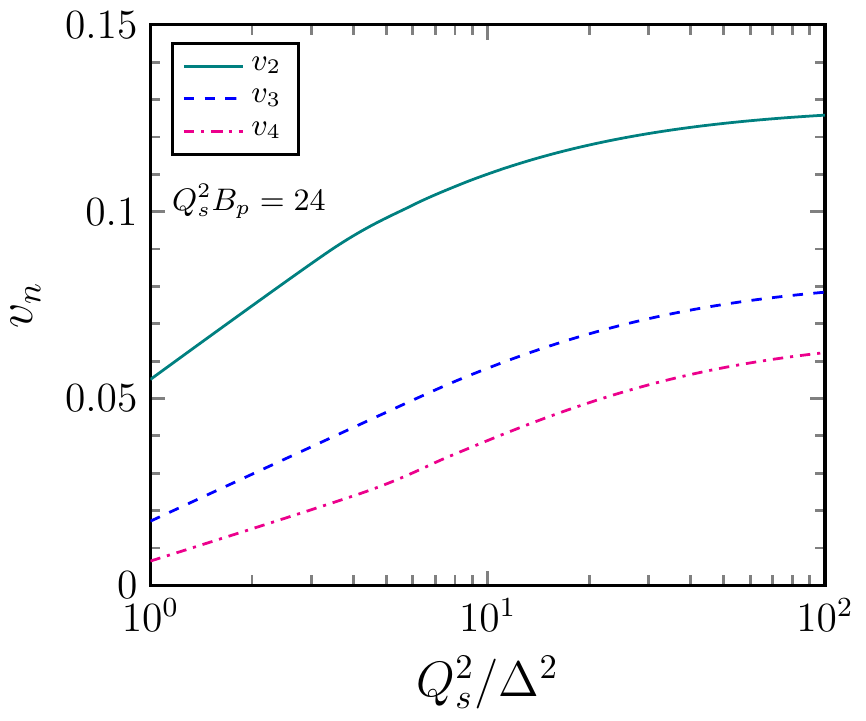}
}
\caption{$v_{2\sim4}\{2\}$ as function of $Q_s^2B_p$ with fixed $\Delta=Q_s/4$ and as function of $Q_s^2/\Delta^2$ with fixed $Q_s^2 B_p=24$. }
\label{vn2}
\end{figure}

In Fig.~\ref{vn2}, we show $v_{2\sim4}\{2\}$ as function of $Q_s^2B_p$ and $Q_s^2/\Delta^2$, which demonstrate the quark-quark channel can produce significant amount of anisotropic harmonics. It is also interesting to note that those $p_\perp$-integrated Fourier coefficients decrease with increasing projectile transverse area $B_p$ or increasing average transverse momentum $\Delta$ coming from the projectile.

At last, we can also obtain the $p_\perp$-dependent Fourier harmonics by integrating over only one of the quark transverse momenta in Eq. (\ref{k2nrp}) and Eq. (\ref{k0}), while keeping the other fixed. Therefore, the $p_\perp$-dependent Fourier harmonics can be defined and computed as 
\begin{eqnarray}
v_{n}\{2\}(p_{\perp})
\equiv
\left( \frac{ \int \rmd^2\bp_2 e^{ in\Delta\phi_{12} } \frac{\rmd^2N}{\rmd^2\bp_{\perp}\rmd^2\bp_2} }{ \int\rmd^2\bp_2\frac{\rmd^2 N}{\rmd^2\bp_{\perp} \rmd^2\bp_2} } \right)^{\frac{1}{2}}
\equiv
\left(\frac{1}{2\pi}\frac{\rmd \kappa_{n}\{2\}(p_{\perp})}{p_{\perp}\rmd p_{\perp}} \Bigg/ \frac{1}{2\pi}\frac{\rmd \kappa_0\{2\}(p_{\perp})}{p_{\perp}\rmd p_{\perp}}\right)^{\frac{1}{2}}.
\end{eqnarray}
%Integrating over $\bp_2$ by Eq. (\ref{intr}), the denominator gives
where denominator is 
\begin{eqnarray}
\frac{\rmd \kappa_0\{2\}(p_{\perp})}{\rmd p_{\perp}}
=
\frac{p_{\perp}}{2b_q}e^{\frac{-p_{\perp}^2}{4b_q}},\ \ \ b_q=\frac{\Delta^2+Q_s^2}{4},
\end{eqnarray}
and the numerator is given by 
\begin{eqnarray}
\frac{\rmd \kappa_{2n}\{2\}(p_{\perp})}{\rmd p_{\perp}}
&=&
\frac{4}{N_c^2} \int_0^1 \rmd\xi \int_0^{\xi} \rmd\eta \frac{1}{(1+\eta Q_s^2B_p)\eta^2 (2n-1)!} \left(\frac{p_{\perp}^2}{4a_q}\right)^{n+1} \frac{1}{p_{\perp}} \sum_{m=n-1}^\infty \frac{m!(2m+2)!}{(2m)!(m-n+1)!} \nonumber
\\
&&\times(\frac{\eta Q_s^2}{8a_q})^{2m+2}\sideset{_1}{_1}{\mathop{F}}(m+n+2;2n+1;\frac{-p_{\perp}^2}{4a_q}),\\
%\end{eqnarray}
%\begin{eqnarray}
\frac{\rmd \kappa_{2n+1}\{2\}(p_{\perp})}{\rmd p_{\perp}}
&=&
\frac{4}{N_c^2} \int_0^1 \rmd\xi \int_0^{\xi} \rmd\eta \frac{1}{(1+\eta Q_s^2B_p)\eta^2 (2n)!} \left(\frac{p_{\perp}^2}{4a_q}\right)^{n+\frac{3}{2}} \frac{1}{p_{\perp}} \sum_{m=n-1}^\infty \frac{(m+\frac{1}{2})!(2m+3)!}{(2m+1)!(m-n+1)!} \nonumber
\\
&&\times(\frac{\eta Q_s^2}{8a_q})^{2m+3}\sideset{_1}{_1}{\mathop{F}}(m+n+3;2n+2;\frac{-p_{\perp}^2}{4a_q}), 
\end{eqnarray}
where $\sideset{_1}{_1}{\mathop{F}}$ represents the generalized hypergeometric function. The numerical results can be found in the next section when we comment on the phenomenological applications of these formulas. 

\subsection{Correlations from the quark-gluon and gluon-gluon channels}

In the case of quark-gluon channel, the relevant quark-gluon-dipole correlator can be defined as
\begin{eqnarray}
\left<D_g(\bx_1,\bx_2)D(\bx_3,\bx_4)\right>
\equiv
\frac{1}{(N_c^2-1)N_c} \left< U(\bx_1)^{ab}U(\bx_2)^{\dagger ab} \text{tr} \left[ V(\bx_3)V(\bx_4)^{\dagger} \right] \right>,\label{DgD}
\end{eqnarray}
where the Wilson line in the adjoint representation $U(\bx)^{ab}$ can be converted into the fundamental representation with the help of the identity $U(\bx)^{ab}\equiv2\text{tr}\left[t^aV(\bx)t^bV(\bx)^{\dagger}\right]$. By using Fierz identity, Eq. (\ref{DgD}) can be cast into
\begin{eqnarray}
\left<D_g(\bx_1,\bx_2)D(\bx_3,\bx_4)\right>
&=&\frac{1}{(N_c^2-1)N_c} \left< \text{tr} \left[ V(\bx_1)V(\bx_2)^{\dagger} \right] \text{tr} \left[ V(\bx_2)V(\bx_1)^{\dagger} \right] \text{tr} \left[ V(\bx_3)V(\bx_4)^{\dagger} \right]\!-\!\text{tr} \left[ V(\bx_3)V(\bx_4)^{\dagger} \right] \right>\nonumber
\\
&=&\frac{N_c^2}{N_c^2-1}\left< D(\bx_1,\bx_2)D(\bx_2,\bx_1)D(\bx_3,\bx_4) - \frac{1}{N_c^2} D(\bx_3,\bx_4) \right>,
\end{eqnarray}
which consists of the expectation value of a 3-dipole correlator plus a 1-dipole correlator. By using the similar procedure developed in the CGC framework \cite{Gelis:2001da,Fujii:2002vh,Blaizot:2004wv,Dominguez:2008aa,Marquet:2010cf,Dominguez:2012ad,Shi:2017gcq}
and following the steps illustrated in Appendix.~\ref{qgc}, one can obtain the following expression for the 3-dipole correlator in the large-$N_c$ limit
\begin{eqnarray}
&&\left<D_g(\bx_1,\bx_2)D(\bx_3,\bx_4)\right> \Bigg|_{ \text{to the} \frac{1}{N_c^2} \text{order} }\nonumber
\\
&=&e^{ -\frac{Q_s^2}{4} (2r_1^2+r_2^2) } \Bigg\{ 1+\frac{1}{N_c^2}
+\frac{ ( \frac{Q_s^2}{2} \br_1\cdot\br_2 )^2 }{N_c^2} \int_0^1 \rmd\xi \int_0^{\xi} \rmd\eta e^{\frac{\eta Q_s^2}{8}[(\br_1-\br_2)^2-4(\bb_1-\bb_2)^2]}\nonumber
\\
&&+\frac{ ( \frac{Q_s^2}{2} \br_1\cdot\br_2 )^2 }{N_c^2} \int_0^1 \rmd\xi \int_0^{\xi} \rmd\eta e^{\frac{\eta Q_s^2}{8}[(\br_1+\br_2)^2-4(\bb_1-\bb_2)^2]}
+\frac{ ( \frac{Q_s^2}{2} r_1^2 )^2 }{N_c^2} \int_0^1 \rmd\xi \int_0^{\xi} \rmd\eta e^{\frac{\eta Q_s^2}{2} r_1^2} \Bigg\}
-\frac{1}{N_c^2}e^{-\frac{Q_s^2}{4}r_2^2}.\label{DgDF}
\end{eqnarray}
It is very interesting to note that the above expression is symmetric under the change $\br_1 \to -\br_1$ (or $\br_2 \to -\br_2$), and thus there is no odd power of $(\br_1\cdot\br_2)$ in $\left<D_g(\bx_1,\bx_2)D(\bx_3,\bx_4)\right>$. According to what we have shown above for the quark-quark channel, this implies that the odd Fourier harmonics vanishes in the quark-gluon channel. 

To compute the Fourier harmonics from the gluon-quark-dipole correlator, it is clear that only the $3^{\text{rd}}$ and $4^{\text{th}}$ terms inside the big brackets of $\left<D_gD\right>$ in Eq. (\ref{DgDF}) contribute and they give
\begin{eqnarray}
\kappa_{n}\{gq\}
&=&\prod_{i=1}^2 \int\frac{ \rmd^2\bb_i \rmd^2\br_i \rmd^2\bp_i } {4\pi^3B_p} e^{ -\frac{b_i^2}{B_p} - \frac{\Delta^2r_i^2}{4} + i\bp_i\cdot\br_i } e^{ -\frac{Q_s^2}
{4} (2r_1^2+r_2^2) } \frac{ (\frac{Q_s^2}{2} \br_1\cdot\br_2)^2 }{N_c^2} \int_0^1 \rmd\xi \int_0^{\xi} \rmd\eta e^{ \frac{\eta Q_s^2}{8} ( r_1^2 + r_2^2 - 4\Delta\bb_{12}^2 ) }\nonumber
\\
&&\times\left( e^{-\frac{Q_s^2}{4} \br_1\cdot\br_2 } + e^{ \frac{Q_s^2}{4} \br_1\cdot\br_2 } \right)e^{ in\Delta\phi_{12} },
\end{eqnarray}
which again clearly show that only even powers of $\br_1\cdot\br_2$ exist. Therefore, similar derivation gives rise to the following integrated even harmonics
\begin{eqnarray}
\kappa_{2n}\{gq\}
=
\frac{8}{N_c^2} \int_0^1 \rmd\xi \int_0^{\xi} \rmd\eta \frac{1}{1+\eta Q_s^2B_p} \frac{n^2}{\eta^2} \sum_{m=n-1}^\infty \frac{ (m!)^2 \binom{2m+2}{m+n+1} }{(2m)!} \left[ \frac{\left(\frac{\eta Q_s^2}{8}\right)^2}{a_g a_q} \right]^{m+1},\ \ \ a_g=\frac{\Delta^2+2Q_s^2-\frac{\eta}{2} Q_s^2}{4},\label{kngq}
\end{eqnarray}
and the $p_\perp$-dependent Fourier even harmonics
\begin{eqnarray}
\frac{\rmd \kappa_0\{gq\}(p_{g,q\perp})}{\rmd p_{g,q\perp}}
&=&
\frac{p_{g,q\perp}}{2b_{g,q}}e^{\frac{-p_{g,q\perp}^2}{4b_{g,q}}},\ \ \ b_g=\frac{\Delta^2+2Q_s^2}{4}, \\
%\end{eqnarray}
%\begin{eqnarray}
\frac{\rmd \kappa_{2n}\{gq\}(p_{g,q\perp})}{\rmd p_{g,q\perp}}
&=&
\frac{8}{N_c^2} \int_0^1 \rmd\xi \int_0^{\xi} \rmd\eta \frac{1}{(1+\eta Q_s^2B_p)\eta^2 (2n-1)!} \left(\frac{p_{g,q\perp}^2}{4a_{g,q}}\right)^{n+1} \frac{1}{p_{g,q\perp}} \sum_{m=n-1}^\infty \frac{m!(2m+2)!}{(2m)!(m-n+1)!} \nonumber
\\
&&\times\left[\frac{(\frac{\eta Q_s^2}{8})^2}{a_g a_q}\right]^{m+1}\sideset{_1}{_1}{\mathop{F}}(m+n+2;2n+1;\frac{-p_{g,q\perp}^2}{4a_{g,q}}).
\end{eqnarray}
In fact, as indicated in the above formulas, depending on whether the momentum of the quark or the gluon is kept fixed, we can obtain the $qg$ channel or the $gq$ channel, respectively. 

In the case of two incoming gluons from the proton projectile, it is easy to see that the relevant gluon-gluon-dipole correlator reads
\begin{eqnarray}
\left<D_g(\bx_1,\bx_2)D_g(\bx_3,\bx_4)\right>
\equiv
\frac{1}{(N_c^2-1)^2}\left< U(\bx_1)^{ab}U(\bx_2)^{\dagger ba} U(\bx_3)^{cd}U(\bx_4)^{\dagger dc} \right>,\label{DgDg}
\end{eqnarray}
which can be written as follows using the similar derivation illustrated in Appendix.~\ref{qgc}.
\begin{eqnarray}
\left<D_{g12}D_{g34}\right>\Bigg|_{\text{to the}\frac{1}{N_c^2}\text{order}}
&=&e^{-\frac{Q_s^2}{2}(r_1^2+r_2^2)}\Bigg\{
1 +\frac{2}{N_c^2}+ \frac{ 2( \frac{Q_s^2}{2} \br_1\cdot\br_2 )^2 }{N_c^2} \int_0^1 \rmd\xi \int_0^{\xi} \rmd\eta e^{ \frac{\eta Q_s^2}{8} [(\br_1-\br_2)^2-4(\bb_1-\bb_2)^2] }\nonumber
\\
&&+ \frac{ 2( \frac{Q_s^2}{2} \br_1\cdot\br_2 )^2 }{N_c^2} \int_0^1 \rmd\xi \int_0^{\xi} \rmd\eta e^{ \frac{\eta Q_s^2}{8} [(\br_1+\br_2)^2-4(\bb_1-\bb_2)^2] }
+ \frac{ ( \frac{Q_s^2}{2} r_1^2 )^2 }{N_c^2} \int_0^1 \rmd\xi \int_0^{\xi} \rmd\eta e^{ \frac{\eta Q_s^2}{2} r_1^2 }\nonumber
\\
&&+\frac{( \frac{Q_s^2}{2} r_2^2 )^2 }{N_c^2} \int_0^1 \rmd\xi \int_0^{\xi} \rmd\eta e^{ \frac{\eta Q_s^2}{2} r_2^2 } \Bigg\}
-\frac{1}{N_c^2}\left( e^{ -\frac{Q_s^2}{2} r_1^2 } + e^{ -\frac{Q_s^2}{2} r_2^2 } \right). \label{ggc}
\end{eqnarray}
The above expression is equivalent to the result found in Ref.~\cite{Kovchegov:2012nd}. In the same way, the Fourier harmonics from two gluon channel can be computed by using the following formula
\begin{eqnarray}
\kappa_{n}\{gg\}
&=&\prod_{i=1}^2 \int \frac{ \rmd^2\bb_i \rmd^2\br_i \rmd^2\bp_i }{4\pi^3B_p} e^{ -\frac{b_i^2}{B_p} - \frac{\Delta^2r_i^2}{4} + i\bp_i\cdot\br_i }
e^{ -\frac{Q_s^2}{2} (r_1^2+r_2^2) } \frac{ 2(\frac{Q_s^2}{2} \br_1\cdot\br_2)^2 }{N_c^2} \int_0^1 \rmd\xi \int_0^{\xi} \rmd\eta e^{ \frac{\eta Q_s^2}{8} ( r_1^2 + r_2^2 - 4\Delta\bb_{12}^2 ) }\nonumber
\\
&&\times\left( e^{ -\frac{Q_s^2}{4} \br_1\cdot\br_2 } + e^{ \frac{Q_s^2}{4} \br_1\cdot\br_2 } \right) e^{ in\Delta\phi_{12} },
\end{eqnarray}
which gives the integrated even harmonics
\begin{eqnarray}
\kappa_{2n}\{gg\}\
=
\frac{16}{N_c^2} \int_0^1 \rmd\xi \int_0^{\xi} \rmd\eta \frac{1}{1+\eta Q_s^2B_p} \frac{n^2}{\eta^2} \sum_{m=n-1}^\infty \frac{ (m!)^2 \binom{2m+2}{m+n+1} }{(2m)!} \left( \frac{\eta Q_s^2}{8 a_g} \right)^{2m+2}.\label{kngg}
\end{eqnarray}
and the $p_\perp$ dependent ones
\begin{eqnarray}
\frac{\rmd \kappa_0\{gg\}(p_{\perp})}{\rmd p_{\perp}}
&=&
\frac{p_{\perp}}{2b_{g}}e^{\frac{-p_{\perp}^2}{4b_{g}}}, \\
%\end{eqnarray}
%\begin{eqnarray}
\frac{\rmd \kappa_{2n}\{gg\}(p_{\perp})}{\rmd p_{\perp}}
&=&
\frac{16}{N_c^2} \int_0^1 \rmd\xi \int_0^{\xi} \rmd\eta \frac{1}{(1+\eta Q_s^2B_p)\eta^2 (2n-1)!} \left(\frac{p_{\perp}^2}{4a_{g}}\right)^{n+1} \frac{1}{p_{\perp}} \sum_{m=n-1}^\infty \frac{m!(2m+2)!}{(2m)!(m-n+1)!} \nonumber
\\
&&\times\left( \frac{\eta Q_s^2}{8 a_g} \right)^{2m+2}\sideset{_1}{_1}{\mathop{F}}(m+n+2;2n+1;\frac{-p_{\perp}^2}{4a_g}).
\end{eqnarray}

\section{Conclusion and outlook}

\begin{figure}[t]
\vskip0.0\linewidth
\centerline{
\includegraphics[width = 0.65\linewidth]{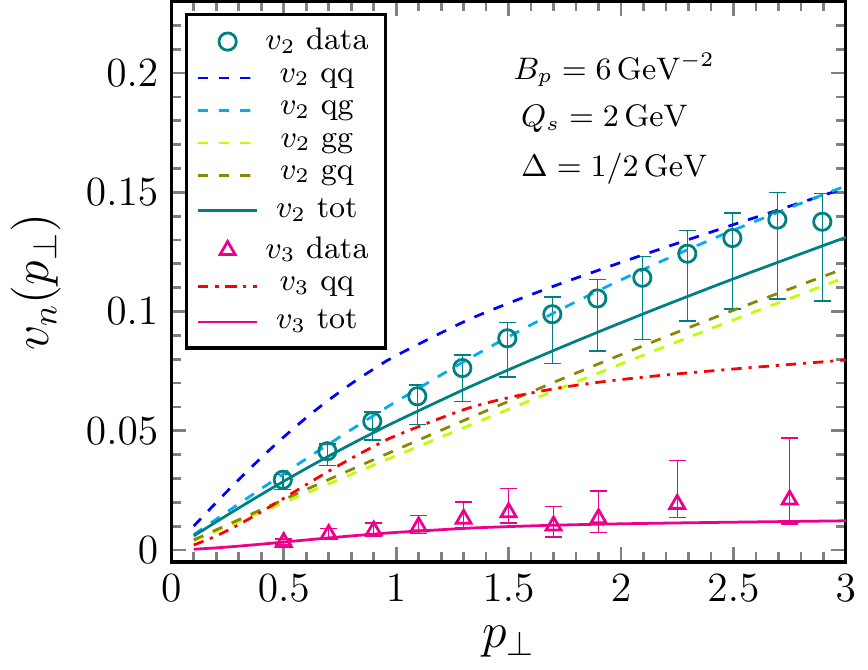} %\includegraphics[width = 0.45\linewidth]{v32tot.pdf}
}
\caption{The azimuthal harmonics $v_{2}\{2\}$ and $v_{3}\{2\}$ as a function of $p_\perp$, compared to the $pAu$ PHENIX data. The separate contributions of the various channels are also shown.}
\label{vnpt2}
\end{figure}

In this section, we comment on the phenomenological implication our the simple Wilson line approach to azimuthal Fourier harmonics. Let us take the recent high-multiplicity $pAu$ data, which are shown in the left plot of Fig.~\ref{vnpt2}, measured by PHENIX collaboration\cite{Aidala:2018mcw} as an example. Guided by previous phenomenological studies, we choose the values $B_p=6\ \textrm{GeV}^{-2}$, $\Delta =0.5\ \textrm{GeV} $ and $Q_s=2\ \textrm{GeV}$, based on our estimates for the RHIC kinematics. It shows that all four channels ($qq$, $qg$, $gq$ and $gg$) roughly give a similar magnitude for $v_2(p_\perp)$, when normalized by their own $\kappa_0$.

To sum up all channels, one needs to take into account the ratio between the gluon density and the quark density, and weight these channels accordingly. By defining $r(x)\equiv g(x)/q(x)$ as the ratio between the gluon density and the total quark density, we can write the total $v_n$ as 
\be
v_{2n}^{\textrm{tot}}= \sqrt{\frac{\kappa_{2n}\{qq\}+r \kappa_{2n}\{qg\}+r\kappa_{2n}\{gq\}+r^2\kappa_{2n}\{gg\}}{\kappa_0\{qq\}+r \kappa_0\{qg\}+r\kappa_0\{gq\}+r^2\kappa_0\{gg\}}},
\ee
for even harmonics. As to the odd harmonics, due to the cancellation between quarks and antiquarks as mentioned earlier, only valence quarks contribute, therefore for $v_3^{\textrm{tot}}$ one has 
\be
v_{3}^{\textrm{tot}}= \frac{r}{r_v}\sqrt{\frac{\kappa_3\{qq\}}{\kappa_0\{qq\}+r \kappa_0\{qg\}+r\kappa_0\{gq\}+r^2\kappa_0\{gg\}}},
\ee 
where $r_v\equiv\ g(x)/q_v(x)$ is the ratio between the gluon density and the total valence quark density. To simplify the calculation, we simply choose $r_v=4$ and $r=1.6$, since $g(x) \simeq 4 q_v(x)$ and $g(x) \simeq 1.6 q(x)$ at $x=0.04$. As shown in Fig.~\ref{vnpt2}, the shape and magnitude of the total $v_2(p_\perp)$, which is computed from the two-particle cumulant method, agree with the data measured using the event plane method. %Presumably, the difference, namely the non-flow contribution, is small.

More interestingly, as discussed in previous sections, the $qq$ channel yields significant $v_3(p_\perp)$, while the other three channels involving gluons Wilson lines give vanishing $v_3(p_\perp)$. We find that the total $v_3(p_\perp)$ is suppressed roughly by a factor of $(1+r)r_v/r\simeq 6.5$ as compared to the quark-quark channel one. Nevertheless, the total $v_3(p_\perp)$ is roughly within the range of PHENIX data. 

%In addition, we expect that a factor of $2\sim 3$ increase as shown in the right plot of Fig.~\ref{vnpt2} for the ratio of $v_3/v_2$ when quark distributions are much larger than gluon distributions. When $x$ is sufficiently large in the proton projectile, we expect that all contributions from gluons, including the three gluon contribution to $v_3$\cite{Kovchegov:2018jun}, should be small. Therefore, it could be interesting to test this prediction by going to forward rapidity region, where quark distributions become dominant, and see if there is an enhancement of $v_3$ or not. 

Admittedly, we do not wish to claim that our simple model can describe the RHIC data perfectly, since we have employed a lot of approximations in the course of our derivation. For example, one can put in the full kinematics and fragmentation functions as well as finite transverse momentum cuts, etc. We also have not included the so-called HBT and Bose enhancement contributions to even harmonics, which require more refined modelling of the proton structure. It would be interesting to see whether one can improved the double Wigner distribution used in our model in order to included those, as well as to compute the harmonics in $dAu$ and $^3$He$Au$ collisions. We will leave these for a future study. Finally, for LHC kinematics\cite{ATLAS:2018jht} where the ratio between the gluon distribution and the valence quark distribution becomes huge, gluon $v_3$ due to higher-order terms (both in $\alpha_s$ and in the projectile density) should also be considered \cite{Kovner:2016jfp,Kovchegov:2018jun,Mace:2018yvl}.

\begin{acknowledgments}
We thank T. Altinoluk, Y. Kovchegov, A. Mueller, S. Munier, R. Venugopalan and F. Yuan for useful discussions and comments. This material is based on the work supported by the Natural Science Foundation of China (NSFC) under Grant Nos.~11575070 and 11435004. The work of CM was supported in part by the Agence Nationale de la Recherche under the project ANR-16-CE31-0019-02. 
\end{acknowledgments}

\appendix
\section{Several Useful Integral Identities}
In order to analytically study the azimuthal harmonics, we employ the following formulas to help us perform various integrations analytically. For azimuthal angular integration of $p_i$ of measured particle, we can use
\begin{eqnarray}
&&\int d\phi_1 d\phi_2 e^{i\bp_1\cdot \br_1+i\bp_2\cdot \br_2}\text{cos}[n(\phi_1-\phi_2)]=(2\pi)^2 J_n(p_1r_1)J_n(p_2r_2) (-1)^n \text{cos}[n(\theta_1-\theta_2)]. \label{intphi}
\end{eqnarray}
For impact parameter integrations, one can use
\begin{eqnarray}
&&\int d^2\bb_1 d^2\bb_2 e^{-\frac{b_1^2+b_2^2}{B_p}-\frac{\eta Q_s^2}{2}(\bb_1-\bb_2)^2}=\int d^2\bb_+ d^2\bb_- e^{-\frac{2\bb_+^2}{B_p}- \frac{\bb_-^2}{2B_p}-\frac{\eta Q_s^2}{2}\bb_-^2} = \frac{\pi^2B_p^2}{1+\eta Q_s^2B_p},\label{intb}
\end{eqnarray}
where $\bb_{+}\equiv\frac{\bb_1+\bb_2}{2}$, $\bb_{-}\equiv\bb_1-\bb_2$. For $\theta_i$ integrals, the following identities are quite useful,
\begin{eqnarray}
&&\int_0^{2\pi} d\theta_1 d\theta_2 \text{cos}[2n(\theta_1-\theta_2)] \text{cos}^{2m}(\theta_1-\theta_2)=\frac{(2\pi)^2}{2^{2m}}\binom{2m}{m+n}, \quad \textrm{with} \, m\geq n, \nonumber
\\
&&\int_0^{2\pi} d\theta_1 d\theta_2 \text{cos}[(2n+1)(\theta_1-\theta_2)] \text{cos}^{2m+1}(\theta_1-\theta_2)=\frac{(2\pi)^2}{2^{2m+1}}\binom{2m+1}{m+n+1}, \quad \textrm{with} \, m\geq n, \label{inttheta}
\end{eqnarray}
where $\binom{2m}{m+n}$ denotes binomial coefficient and it should vanish when $m< n$ in our notation. In the end, for the integration over the dipole size and its conjugating momentum, we can employ 
%\begin{eqnarray}
%&&\int_0^\infty rdr \int_0^\infty pdp \ e^{-ar^2} J_{2n}(pr) r^{2m+2}=\frac{d}{d^m(-a)}\int_0^\infty rdr \int_0^\infty pdp \ e^{-ar^2} J_{2n}(pr) \ r^2 \notag \\
%&&=\frac{d}{d^m(-a)}\frac{n}{a}=\frac{m!n}{a^{m+1}},\nonumber\\
%&&\int_0^\infty rdr \int_0^\infty pdp  e^{-ar^2} J_{2n+1}(pr) r^{2m+3}=\frac{d}{d^m(-a)}\int_0^\infty rdr  \int_0^\infty pdp  e^{-ar^2} J_{2n+1}(pr) r^3 \notag \\
%&&=\frac{d}{d^m(-a)}\frac{\sqrt{\pi}(2n+1)}{4a^\frac{3}{2}}=\frac{\sqrt{\pi}(2n+1)}{4a^{m+\frac{3}{2}}}\frac{(2m+1)!!}{2^m}. %\nonumber\label{intrp}
%\end{eqnarray}
\begin{eqnarray}
&&\int_0^\infty r\rmd r \int_0^\infty p\rmd p e^{-ar^2} J_n(pr) r^m
=
\frac{ \Gamma(\frac{m}{2}) \frac{n}{2} }{a^{\frac{m}{2}}}, \label{intrp}
\\
&&\int_0^\infty p\rmd p e^{-ar^2} J_n(pr) r^m
=
\frac{ p^n \Gamma(\frac{m+n}{2}+1) }{ 2^{n+1} n! a^{\frac{m+n}{2}+1} } \sideset{_1}{_1}{\mathop{F}} (\frac{m+n}{2}+1;n+1;\frac{-p^2}{4a}).\label{intr}
\end{eqnarray}

\section{The detailed evaluation of quark-gluon-dipole amplitude}\label{qgc}
Here we show detailed steps on the computation of the quark-gluon-dipole amplitude. The 3-dipole correlator in the CGC formalism\cite{Gelis:2001da, Fujii:2002vh,Blaizot:2004wv, Dominguez:2008aa,Marquet:2010cf,Dominguez:2012ad, Shi:2017gcq} can be written as
\begin{eqnarray}
\left<D(\bx_1,\bx_2)D(\bx_2,\bx_1)D(\bx_3,\bx_4)\right>
=
\frac{T_{\text{3-dipole}}}{N_c^3}
\left(\begin{matrix} N_c^3 & N_{c(1\times3)}^2 & N_{c(1\times2)} \end{matrix}\right)
e^{M_{\text{3-dipole}}}
\left(\begin{matrix} 1 \\ 0_{(5\times1)} \end{matrix}\right),
\end{eqnarray}
where $M_{\text{3-dipole}}$ is the corresponding color transition matrix. Up to the $\frac{1}{N_c^2}$ order, the 3-dipole correlator is
\begin{eqnarray}
\left<D(\bx_1,\bx_2)D(\bx_2,\bx_1)D(\bx_3,\bx_4)\right> \Bigg|_{\text{to the} \frac{1}{N_c^2} \text{order}}
=
\frac{T_{\text{3-dipole}}}{N_c^3}
\left(\begin{matrix} N_c^3 & N_{c(1\times3)}^2 \end{matrix}\right)
e^{M'_{\text{3-dipole}}}\left(\begin{matrix} 1 \\ 0_{(3\times1)} \end{matrix}\right),
\end{eqnarray}
where $T_\text{3-dipole}=e^{-\frac{C_F}{2} \mu^2 \sum\limits_{i=1}^{6} L_{ii}}$ is the so-called tadpole contribution and $M'_{\text{3-dipole}}$ is the $4\times4$ submatrix of $M_{\text{3-dipole}}$ which takes the form
\begin{equation}
M'_{\text{3-dipole}}
=
\mu^2\left(\begin{matrix}
LF_{12,34,21} & \frac{1}{2}F_{1423} & 0 & \frac{1}{2}F_{3142}
\\
\frac{1}{2}F_{1243} & LF_{14,32,21} & 0 & 0
\\
\frac{1}{2}F_{1212} & 0 & LF_{11,34,22} & 0
\\
\frac{1}{2}F_{3412} & 0 & 0 & LF_{12,31,24}
\end{matrix}\right)
\equiv
\left(\begin{matrix}
M_{1(1\times1)} & M_{4(1\times3)} \\ M_{2(3\times1)} & M_{3(3\times3)} \end{matrix}\right),
\end{equation}
where $LF_{ab,cd,ef}\equiv C_F(L_{ab}+L_{cd}+L_{ef})+\frac{1}{2N_c}(F_{abcd}+F_{abef}+F_{cdef})$ and 
\begin{eqnarray}
\Gamma_{ij}\equiv \mu^2 \left(L_{ii}+L_{jj}-2 L_{ij}\right) = \frac{Q_s^2}{2C_F} (\bx_i- \bx_j)^2,
\ \ \
\mu^2 F_{ijkl} = \frac{Q_s^2}{2C_F} (\bx_i-\bx_j) \cdot (\bx_k-\bx_l).\label{LF}
\end{eqnarray}
Here $Q_s^2$ and $\mu^2$ are related to the density of the target nucleus. In our calculation, $Q_s^2$ is treated as a constant. By simplifying the above matrix expression, we can arrive at
\begin{eqnarray}
&&\left<D(\bx_1,\bx_2)D(\bx_2,\bx_1)D(\bx_3,\bx_4)\right> \Bigg|_{ \text{to the} \frac{1}{N_c^2} \text{order} }\nonumber
\\
&=& e^{-\frac{C_F}{2}\left(\Gamma_{12}+\Gamma_{34}+\Gamma_{21}\right)} \Bigg[ 1 + \left(\frac{\mu^2}{2}F_{1243}\right)^2 \int_0^1 \rmd\xi \int_0^{\xi} \rmd\eta e^{\eta N_c \frac{\mu^2}{2}F_{1342}}
 + \left( \frac{\mu^2}{2} F_{3412} \right)^2 \int_0^1 \rmd\xi \int_0^{\xi} \rmd\eta e^{ \eta N_c \frac{\mu^2}{2} F_{3214} }\nonumber
\\
&&+\left( \frac{\mu^2}{2} F_{1212} \right)^2 \int_0^1 \rmd\xi \int_0^{\xi} \rmd\eta e^{ \eta N_c \frac{\mu^2}{2} F_{1212} } \Bigg].\label{DDD}
\end{eqnarray}
Substituting Eq. (\ref{LF}) in Eq. (\ref{DDD}) and together with $\left<D\left(\bx_3,\bx_4\right)\right>=e^{-\frac{Q_s^2}{4}r_2^2}$, we can obtain the gluon-quark dipole correlator in Eq. (\ref{DgDF}). %T_{\text{1-dipole}} T_{\text{3-dipole}} e^{ \mu^2 C_F L_{12,34,21} }

\end{document}